\documentstyle[12pt]{article}
\begin{document}
\pagestyle{plain}

\newcommand{\be}{\begin{equation}}
\newcommand{\ee}{\end{equation}}
\newcommand{\bea}{\begin{eqnarray}}
\newcommand{\eea}{\end{eqnarray}}
\newcommand{\vp}{\varphi}
\newcommand{\pr}{\prime}
\newcommand{\sech} {{\rm sech}}
\newcommand{\cosech} {{\rm cosech}}
\newcommand{\psib} {\bar{\psi}}
\newcommand{\cosec} {{\rm cosec}}

\title{Truncated Harmonic Osillator and Parasupersymmetric
Quantum Mechanics}
\author{
 B. Bagchi \\
{\small \sl Department of Applied Mathematics, University of Calcutta,}\\
{\small \sl 92 Acharya P.C.Road, Calcutta 700 009, India}\\
{\small \sl  }\\
\ S.N. Biswas \\
{\small \sl  Department of Physics, Delhi University,} \\
{\small \sl Delhi - 110 007, India}
{\small \sl }\\
\ Avinash Khare \\
{\small \sl Institute of Physics, Sachivalaya Marg,} \\
{\small \sl Bhubaneswar 751 005, India} \\
\and P.K. Roy  \\
{\small \sl Department of Physics, Haldia Government College,} \\
{\small \sl Haldia 721 657, India.}}

\maketitle

\begin{abstract}
We discuss in detail the parasupersymmetric quantum mechanics of arbitrary
order where the parasupersymmetry is between the normal bosons and those
corresponding to the truncated harmonic oscillator. We show that even though
the parasusy algebra is different from that of the usual parasusy quantum
mechanics, still the consequences of the two are identical. We further
show that the parasupersymmetric quantum mechanics  of arbitrary order p
can also be rewritten in terms of p supercharges (i.e. all of which
obey $Q_i^{2} = 0$).
However, the Hamiltonian cannot be expressed in a simple form
in terms of the p supercharges except in a special case. A model of
conformal parasupersymmetry is also discussed and it is shown that in this
case, the p supercharges, the p conformal supercharges along with Hamiltonian
H, conformal generator K and dilatation generator D form a closed algebra.
\end{abstract}
\vfill
\eject

A great deal of attention is now being paid to study
\cite{kuang,pegg,peng,wilson}
 quantum
mechanics in a finite dmensional Hilbert space (FHS).
In particular, we would like to mention the recent developments
\cite{pegg,peng,wilson} in quantum
phase theory which deals \cite{pegg} with a quantized harmonic oscillator in
a FHS and which finds interesting applications \cite{wilson} in problems of
quantum optics.

Recently, two of us (BB and PKR) studied \cite{br} some basic properties of
these
oscillators. In particular it was pointed out that the raising and
lowering operators of the truncated oscillator behave like parafermi
osillator. Inspired by this similarity, a parasupersymmetric quantum
mechanics (PSQM) of order 2 was also written down where the parasusy is between
the usual bosons and the truncated bosons. However, the explicit form of the
charge was not written down. Further the consequences were also not elaborated
upon. The purpose of this note is to generalize this construction to arbitrary
order.
In particular, we show that for these  PSQM models
of arbitrary order p, the algebra is given by
\be \label{1}
Q^{p+1} = 0 ; \hspace{.2in} [H, Q] = 0
\ee
\be \label{2}
Q^p Q^{+}+Q^{p-1}Q^{+}Q+...+Q^{+}Q^{p} = p(p+1)Q^{p-1}H
\ee
and the Hermitian conjugated relations  and discuss their consequences in some
detail. In particular, we show that the consequences following from this
algebra
are identical to those following from the well known PSQM
model of the same order p \cite{rs} \cite{ak} even though the two algebras are
different.
In particular,
whereas eq. (\ref{1}) is identical in the two schemes, eq. (\ref{2}) is
different in the two schemes in the sence that in the well known case the
coefficient on the r.h.s. is 2p instead of p(p+1) in eq. (\ref{2}).
In view of the identical consequences, it is worth examining as to why the PSQM
of order p can be written down in an alternative way. To that end we show that
 one can infact express PSQM of order p in terms of p super
(rather than parasuper) charges all of which satisfy
$Q_i^{2}$ =0 and further all of them commute with the Hamiltonian. However,
unlike the usual supersymmetric (SUSY) quantum mechanics (QM), here H cannot be
simply
expressed in terms of the p supercharges except in a special case. In the
special case we show that the Hamiltonian has a very simple expression in
terms of the p supercharges
\be \label{eqi}
Q_{1}Q_{1}^{+}+\sum^{p}_{j=1} Q_{j}^{+}Q_{j} = 2H
\ee
We also discuss a para superconformal model of order p and show that the
dilatation and conformal operators also can similarly be expressed in
quadratic form in terms of the p SUSY and p para superconformal charges.

Let us start with the truncated raising and lowering operatorss
$a^+$ and a. It is well known that if one truncates at (p+1)'th level
(p $>0$ is an integer) then a and $a^+$ can be represented by
$(p+1)\times(p+1)$
matrices and they
satisfy the commutation relation \cite{buc}
\be \label{3}
 [ a, a^+ ] = I - (p+1)K
\ee
 where I is $(p+1)\times(p+1)$ unit matrix while $K = diag(0,0,...,0,1)$
with Ka = 0 and further $K^2 = K\not = 0$. As shown by Kleeman \cite{kle},
the irreducible representations of eq.(\ref{3}) are the same as those for
the scheme
\be \label{4}
[a, a^{+}a] = a;\hspace{.2cm}   a^{p+1} =0 ;  \hspace{.2in}  a^j \not = 0
\hspace{.2in} if \hspace{.1in}j < (p+1).
\ee
A convenient set of representation of the matrices a and $a^+$ is given by
\be \label{5}
(a)_{\alpha\beta} = \sqrt{\alpha}\delta_{\alpha+1,\beta}
\ee
\be \label{5a}
(a^+)_{\alpha,\beta} =\sqrt{\beta}\delta_{\alpha,\beta+1}
\ee
where $\alpha,\beta$ = 1,2,...(p+1). As shown in \cite{br}, the nontrivial
multilinear relation between a and $a^+$ is given by
\be \label{6}
a^{p}a^{+} +a^{p-1}a^{+}a +...+aa^{+}a^{p-1} +a^{+}a^{p} = {p(p+1) \over
2}a^{p-1}.
\ee

These relations are strikingly similar to those of parafermi oscillator of
order
p \cite{ak} except that in the later case, the coefficient on the right hand
side
is p(p+1)(p+2)/6 unlike p(p+1)/2  in eq.(\ref{6}). As expected, for the case
of the Fermi oscillator (p=1), both the coefficients are same while they are
different
otherwise.

Motivated by the nontrivial relation between a and $a^+$ as given by
eq.(\ref{6})
it is worth enquiring if one can construct a kind of PSQM of order p in which
there will be symmetry between bosons and truncated bosons of order p. It turns
out that the answer to the question is yes. In particular, on choosing the
parasusy
charges Q and $Q^{+}$ as $(p+1)\times (p+1)$ matrices as given by
\be \label{7}
(Q)_{\alpha\beta}=b^{+}a=\sqrt{\alpha}(P+iW_{\alpha}) \delta_{\alpha+1,\beta}
\ee
\be \label{8}
(Q^{\dag})_{\alpha\beta}=ba^{+}=\sqrt{\beta}(P-iW_{\beta})
\delta_{\alpha,\beta+1}
\ee
where b, $b^{+}$ denote the bosonic annihilation and creation operators
 and $\alpha,\beta = 1,2,..., (p+1)$, so that $Q$ and $Q^{\dag}$
automatically satisfy $Q^{p+1} = 0 = (Q^{\dag})^{p+1}$ .
Further, it is easily shown that the Hamiltonian $(\hbar = m = 1)$
\be \label{9}
(H)_{\alpha\beta} = H_{\alpha} \delta_{\alpha\beta},
\ee
where $(r = 1,2,..., p)$
\bea \label{10}
H_r = {P^2 \over 2}+{1 \over 2}(W^2_r-W'_r)+{1 \over 2} C_r \nonumber \\
H_{p+1} = {P^2 \over 2}+{1 \over 2}(W^2_p+W'_p)+{1 \over 2} C_p
\eea
commutes with the PARASUSY charges $Q$ and $Q^{\dag}$ (i.e. $[H,Q]=0=[H,Q^+]$)
provided $(s = 2,3..., p)$
\be \label{11}
W^2_{s-1} + W'_{s-1} + C_{s-1} = W^2_s - W'_s + C_s .
\ee
Here $C_1,C_2,...,C_p$ are arbitrary constants with the dimension of
energy. It turns out that the nontrivial relation given by eq.(\ref{2})
between $Q,Q^{\dag}$ and $H$ is satisfied provided
\be \label{13}
C_1 + 2C_2+ ...+pC_p = 0 .
\ee

It is interesting to notice that the PARASUSY charge as well as the algebra as
given by eqs.(\ref{1}),
(\ref{2}), (\ref{7}), (\ref{8}) and (\ref{13}) is very similar to that of
standard PSQM of order p
\cite{ak} except that in the standard case the coefficient on the r.h.s. of
eq.(\ref{2}) is 2p instead of p(p+1)/2 and instead of eq.(\ref{13}) in the
standard case one has
\be \label{14}
C_1+C_2+...+C_p = 0
\ee
Besides, unlike in eq. (\ref{7}), in the standard case, Q is defined
without the factor of $\sqrt\alpha$.
However, the Hamiltonian and the relation between the superpotentials as given
by
eqs.(\ref{9}) to (\ref{11}) are identical in the two cases. As a result the
consequences following from the two different PSQM scemes of order p are
identical. In particular, as shown in \cite{ak}, in both the cases (i)the
spectrum is not necessarily positive semidefinite unlike in SUSY QM (ii) the
spectrum is (p+1)-fold degenerate atleast above the first p levels while the
ground state could be 1,2,...,p fold degenerate depending on the form of the
superpotentials and (iii) one can associate p ordinary SUSY QM Hamiltonians.

Why do the two seemingly different PSQM schemes give the same consequences?
The point is that in the case of parasusy of order p, one has p independent
parasusy charges and in the two schemes
one has merely used two of the p independent forms of Q. It is then clear that
one can infact define p seemingly different PSQM schemes of order p but all
of them will have identical consequences. For example, the parafermi operators
are usually defined by the following $(p+1)\times(p+1)$ matrices \cite{ak}
\be \label{15}
(a)_{\alpha\beta} = \sqrt{\alpha (p-\alpha +1)}\delta_{\alpha+1,\beta}
\ee
\be \label{15a}
(a^+)_{\alpha,\beta} =\sqrt{\beta(p-\beta+1)}\delta_{\alpha,\beta+1}
\ee
So one could as well have defined the parasusy charges by
\be \label{16}
(Q)_{\alpha\beta}=b^{+}a=\sqrt{\alpha(p-\alpha+1)}(P+iW_{\alpha})
\delta_{\alpha+1,\beta}
\ee
\be \label{16a}
(Q^{\dag})_{\alpha\beta}=ba^{+}=\sqrt{\beta(p-\beta+1)}(P-iW_{\beta})
\delta_{\alpha,\beta+1}
\ee
instead of the usual choice without the squareroot factor \cite{ak}. It is
easily
shown that in this case too the parasusy charges Q and $Q^+$ satisfy the
algebra
as given by eqs. (\ref{1}), (\ref{2}) and (\ref{13})  except that the factor on
the
r.h.s. of eq. (\ref{2}) is now p(p+1)(p+2)/3 and the constants $C_i$ satisfy
\be \label{17}
p(C_1+C_p) + 2(p-1)(C_2+C_{p-1})+...+{(p+1) \over 2} C_{p+1 \over 2} =0,
\hspace{.2in} p \hspace{.1in} odd
\ee
\be \label{18}
p(C_1+C_p) + 2(p-1)(C_2+C_{p-1})+...+{p(p+2) \over 4} (C_{{p \over
2}}+C_{{{p+2} \over 2}}) =0, \hspace{.2in} p \hspace{.1in} even
\ee
instead of eq. (\ref{13}). However, as before the Hamiltonian and the relation
between the various superpotentials is unaltered and hence one would get
the same consequences as in the standard PSQM case \cite{ak}.

At this stage it is worth asking if parasusy QM of order p can be put in an
alternative form
by making use of the fact that there are p independent parasupercharges
\cite{ak}?
If yes this would be analogous to the so called Green construction for
parafermi
and parabose operators \cite{gre}.
We now show that the answer to the question is yes.
Let us first note that the supercharge as given by eq. (\ref{7}) can be written
down as a linear combination of the following p supercharges \cite{bb}
\be \label{19}
Q =\sum^{p}_{j = 1}\sqrt{j}Q_{j}  \hspace{.2in} \j = 1,2,...p
\ee
where
\be \label{20}
(Q_j)_{\alpha\beta} = (P-iW_{j})\delta_{\alpha+1,\beta=j+1}
\ee
It is easily checked that these p charges $Q_{j}$ are infact supercharges in
the sense
that all of them satisfy $Q_{j}^2 =0$. Further, all of them commute with the
Hamiltonian as given by eq. (\ref{9}) provided condition (\ref{11}) is
satisfied.
Besides they satisfy
\be \label{21}
Q_{i}Q_{j} = 0 \hspace{.2in} if \hspace{.1in} j \not =  i+1
\ee
\be \label{21a}
 Q_{i}Q_{j}^{+} = 0 = Q_{i}^{+}Q_j  \hspace{.2in} if \hspace{.1in} i \not = j
\ee
However, there is one respect in which these charges are different from the
usual
SUSY charges in that unlike in that case, the nontrivial relation of the usual
parasusy algebra (i.e. eq. (\ref{2}) but with 2p on the r.h.s. instead of
p(p+1))
now contains product of all p charges i.e.
\be \label{22}
Q_{1}[Q_{1}^{+}Q_{1}Q_{2}...Q_{p-1} +Q_{2}Q_{2}^{+}Q_{2}...Q_{p-1}+...+
Q_{2}Q_{3}...Q_{p}Q_{p}^{+}]
= 2pQ_{1}Q_{2}...Q_{p-1}H
\ee
\be \label{23}
[Q_{1}^{+}Q_{1}Q_{2}...Q_{p-1} +Q_{2}Q_{2}^{+}Q_{2}...Q_{p-1}+...+
Q_{2}Q_{3}...Q_{p}Q_{p}^{+}]Q_{p}
= 2pQ_{2}Q_{3}...Q_{p}H
\ee
provided eq. (\ref{14}) is satisfied. If one instead considers other versions
of
PSQM of order p then one would have  similar relations but with different
weight
factors between the various terms and also different relations between $C_{i}$
which can easily
be worked out.

There is one special case however when the algebra takes a particularly simple
form. In particular when all the constants $C_{i}$ are zero then it is easily
checked that
the Hamiltonian can be written as a sum over quadratic pieces in Q as given by
eq. (\ref{eqi})
which is a generalization of the SUSY algebra in the case of p supercharges. In
this case, clearly the spectrum is positive semidefinite
and most of the results
about SUSY breaking etc. would apply. Further all the excited states are
always $(p+1)$-fold  degenerate. It is amusing to note that in ortho
supersymmetricsy QM
too \cite{kmr}
the relation between H and charges is exactly as given by eq. (\ref{eqi}).

 Following the work of \cite{ak}, we now consider a specific PSQM model of
order
p which in addition is conformally invariant and show that the conformal PSQM
algebra is rather simple. Let us consider the choice
\be \label{24}
W_1 =W_2 =....=W_p = -{\lambda \over x}
\ee
Note that in this case the condition (\ref{11}) is trivially satisfied when all
$C_i$ are zero. The interesting point is that in this case, apart from the p
parafermionic charges $Q_i$, we can also define the dilatation operator D, the
conformal operator K and p para superconformal charges $S_j$ so that they form
a closed
algebra. In particular, on defining
\be \label{25}
D =-{1 \over 4} (xP+Px) ; \hspace{.2in} K =x^2/2
\ee
\be \label{26}
(S_j)_{\alpha\beta} = -x\delta_{\alpha+1,\beta=j+1}
\ee
it is easy to show that the algebra satisfied by D, H and K is standard
\be \label{27}
[H, K] = 2iD , \hspace{.2in} [D, K] = iK , \hspace{.2in} [D,H] =-iH.
\ee
Further
\be
[K, S_j] =0 , \hspace{.2in} [H, S_j] = iQ_j , \hspace{.2in} [K, Q_j] =iS_j ,
\ee
\be
[D, Q_j] =-{i \over 2}Q_j , \hspace{.2in} [D, S_j] = {i \over 2}S_j.
\ee
Besides, apart from the parasusy algebra as described above (with $C_i$ = 0),
we have
\be
S_{i}S_{j} =0 =Q_{i}S_{j}=S_{i}Q_{j} \hspace{.2in} if \hspace{.1in} j \not =
i+1
\ee
\be
Q_{i}S_{j}^{+} =0 = S_{i}S_{j}^{+} = Q{i}^{+}S_{j} \hspace{.2in} if
\hspace{.1in} i \not = j
\ee
\be
S_1S_1^{+} + \sum^p_{j=1} S_{j}^{+}S_{j} = 2K
\ee
\be
S_1Q_1^{+} + Q_{1}S_{1}^{+} + \sum^p_{j=1} (S_{j}^{+}Q_j +Q_{j}^{+}S_{j}) = 4D
\ee
It is quite remarkable that an identical algebra also follows in the case of
the conformal ortho supersymmetric case \cite{kmr}.

\pagebreak

\end{document}